# Dark energy: the cost of function in protein evolution

Ezequiel A. Galpern[1], Federico Caamaño[2,3], Ignacio E. Sánchez[2,3] and Diego U. Ferreiro[2,3*]

1-Centre for Genomic Regulation, The Barcelona Institute of Science and Technology, Barcelona 08003, Spain.
2- Protein Physiology Lab, Departamento de Química Biológica, Facultad de Ciencias Exactas y Naturales, Universidad de Buenos Aires, Buenos Aires C1428EGA, Argentina.
3- Instituto de Química Biológica de la Facultad de Ciencias Exactas y Naturales, Consejo Nacional de Investigaciones Científicas y Técnicas - Universidad de Buenos Aires, Buenos Aires C1428EGA, Argentina
*to whom correspondence should be addressed: ferreiro@qb.fcen.uba.ar
—-----
Short title: Dark energy in the protein universe

**Abstract**

The evolutionary fate of proteins is driven by both folding stability and biological function, dual constraints that often conflict, creating frustration and imposing functional costs beyond stability. These costs can be captured by a “dark energy”: the difference between the evolutionary energy of protein sequences and their physical folding energy. Recent advances in deep mutational scanning, protein language models, and inverse-folding models have enabled the quantification of dark energy across the protein universe. We review the computational and experimental approaches that disentangle folding and function at scale, revealing a dark energy component and providing new insights into how biological information flows from sequence to structure to function and back to sequence.

Highlights

- Dark energy quantifies the fundamental tension between protein stability and activity
- About 25% of positions in globular proteins carry significant dark energy.
- Dark energy localizes around active, allosteric and binding sites

## Introduction

Protein molecules are the prime information processing elements in living matter [1]. Composed of linear polar chains of just a handful of amino acids, these large molecules spontaneously fold and perform distinct chemical activities that relate to their biological functions. This functional capacity is not simply a physicochemical property of the folded molecule; it is a genuine biological concept that is emergent from the proteins' structures interacting with their environment, a molecular scale Protein Physiology. Protein folding and function can be pictured as the process of decoding the one-dimensional digital sequence information into their analog three-dimensional structural information: sequence → function [2]. However, the vast majority of the possible polypeptide sequences do not display proteinaceous characteristics as they do not fold nor perform biological functions. Protein sequences are not merely chosen for their ability to fold; they are selected for their ability to perform a task. This introduces a powerful downward causation where function, an emergent property of the folded protein, exerts a top-down influence on the sequence of the protein itself. In other words, the necessity for specific functions creates additional evolutionary constraints that feed back and shape the primary structure: function → sequence, closing a physiological informational loop (Figure 1). Molecular evolution can be viewed as a process where analog information is being digitized into the genomes. It is in this recurrent flow of biological information that conflicting goals meet in the specification of the structures [3].

## Theoretical considerations

The 'principle of minimal frustration', introduced by Bryngelson and Wolynes [4], provides a deep quantitative perspective for studying folding stability and functional constraints in biomolecules. To fold into specific stable three-dimensional structures, amino acid chains need to establish favorable interactions that are strong enough to overcome competing interactions and compensate for the entropy loss associated with chain folding. Foldable natural proteins have a funnel-like landscape with a large energy gap between the correctly folded structure and competing configurations. Because some interactions conflict with folding, the landscape retains some frustration, but this frustration should be low to ensure rapid folding. Quantitatively, in foldable proteins the critical temperature characterizing the folding transition $T_f$, which is directly related to the energy gap, is higher than the glass transition temperature $T_g$, which is related to the energy of the competing non-native states. Searching for local violations of the principle of minimal frustration in natural proteins exploits the limited coding capacity of sequences and highlights functional constraints. Local frustration can be computationally quantified using the change in folding energy of the native structure under configurational or mutational perturbations [5]. Without access to functional labels or explicit evolutionary information, frustration analysis can localize binding and catalytic sites, allosteric residues, and fuzzy regions of proteins [6,7]. In general local frustration is more evolutionarily conserved than the primary structure itself [8,9].

Building on this physical perspective, protein evolution can itself be viewed through the lens of an energy landscape [10] (Figure 1). Just as a folding protein navigates a funnel-shaped landscape toward its native structure, evolving populations traverse a sequence space landscape where fitness ($\Psi^{evo}$) defines the topography [11]. The evolutionary landscape of a protein family is funneled toward the selected sequences, whereas random amino acid chains lie at the top, separated from the selected sequences by a deep gap. Sequences with sufficiently low folding energy are selected at an apparent selection temperature, $T_{sel}$ [12], which is connected to $T_f$ and $T_g$ [13].

The evolutionary landscape of protein physiology includes both folding and functional constraints [14,15]. In principle, it could be decomposed into a folding landscape and a landscape that reflects purely functional constraints [16], which can be called a ‘Dark Energy’ [17]. The biological and informational relationship between these three landscapes is pictured in Figure 1. We will review here evidence for the existence of Dark Energy, methods for its quantification and its relation to protein physiology.

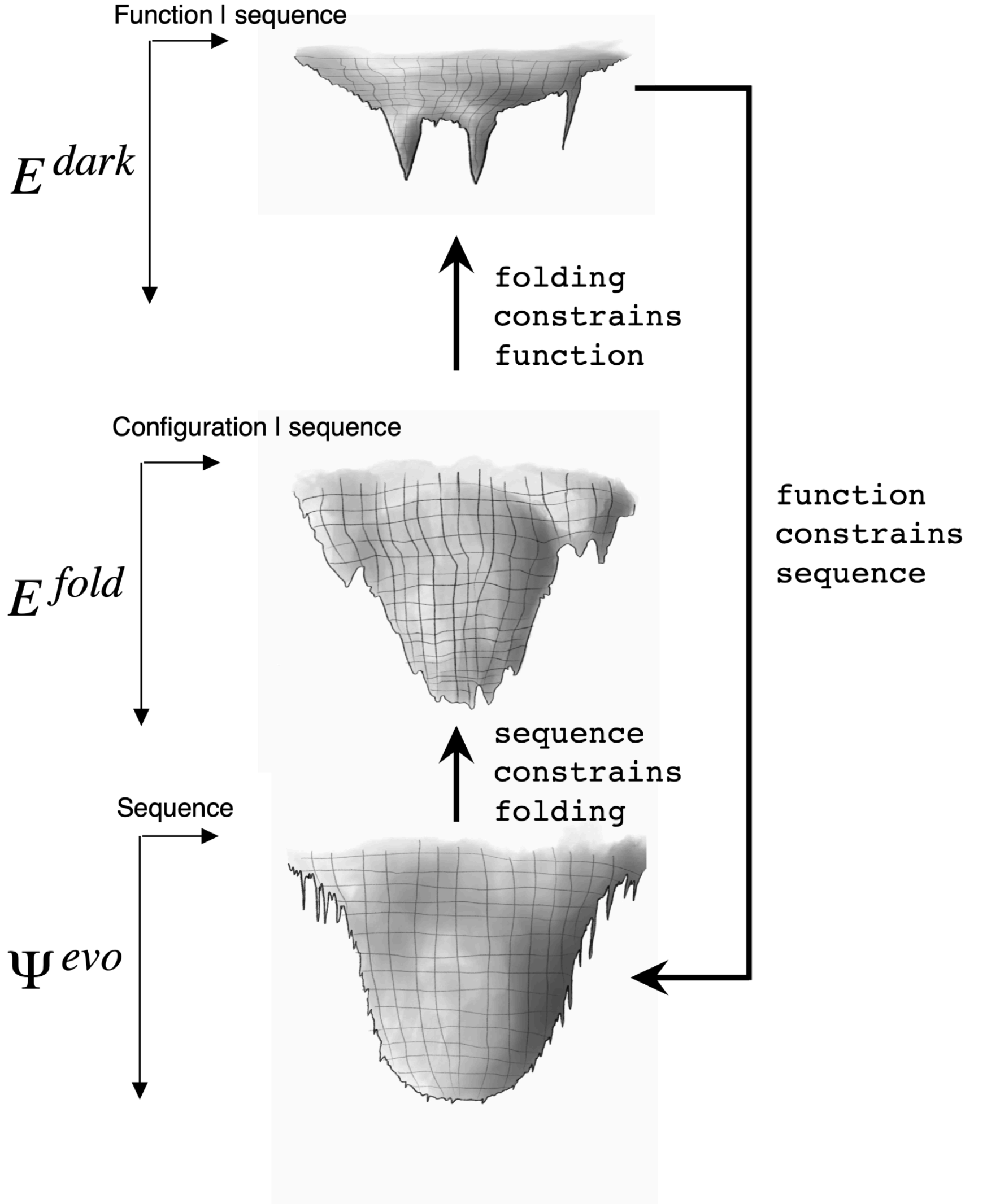

Figure 1: The multiple landscapes of protein physiology. Bottom: Sequences can be pictured to explore a sequence space, whose energy $\Psi^{evo}$ is proportional to their evolutionary likelihood. Middle: the polypeptide explores a configurational space, in a landscape shaped by the folding energy $E^{fold}$. Top: the protein activity can be pictured to occupy a functional landscape, whose energy is proportional to $E^{dark}$. The relationship between the three landscapes and the corresponding energies can be understood in terms of a closed informational loop of emergent biological phenomena. The sequences constrain the structures, which constrain the functions of the proteins, which in turn constrain the sequences.

**Definition of Dark Energy**

If folding were the only physicochemical constraint on molecular evolution, the energy landscape used by nature to select protein sequences should match the one that folds these sequences into functional proteins [18]. However, discrepancies between the evolutionary and folding landscapes are expected as discussed above. Testing this hypothesis requires a quantitative comparison of these two landscapes, which only recently became feasible. Sequence-related stability changes in the folding landscape can be quantified via experiment or computation, for example with the AWSEM coarse-grained potential grounded on energy landscape theory [19]. Within a protein family, a dimensionless potential that quantifies sequence likelihood, sometimes called evolutionary energy, can be inferred from sequence covariation using statistical [20] or deep-learning methods [21]. To a first approximation, such evolutionary energy is nothing more than folding energy rescaled at $T_{sel}$ [18,22], a parameter that can be computed from the variance ratio of the folding and evolutionary energy changes upon point mutations [17,23,24]. In sum, we can measure in the same scale the perturbations caused by local sequence variants to the energetics of the protein folding process and to the corresponding change to the apparent selection landscape of sequences. The difference between evolutionary free energy and the physical folding free energy can be described as the 'dark energy' of the system, [17]

$$E^{dark}(\sigma) = k_B T_{sel}^{fold} \Psi^{evo}(\sigma) - E^{fold}(\sigma) \,. \quad (1)$$

Dark energy summarizes and quantifies the functional constraints on evolution beyond folding-stability (Figure 1). This quantity can be estimated at the variant level (Figure 2). Changes in folding energy are expected to be largest when residues in the protein's structural core are mutated, whereas changes in dark energy peak in and around active sites or binding surfaces, which are typically frustrated regions. When functional constraints can be reduced to a single chemical activity, dark energy can be compared with a physicochemical energy, and a functional selection temperature can be computed [17]. Whereas frustration analysis exploits local violations of the minimal frustration principle to identify functional regions, dark energy quantifies the deviations from the approximation that folding is the sole evolutionary constraint, thus measuring the cost of function to the folding landscape.

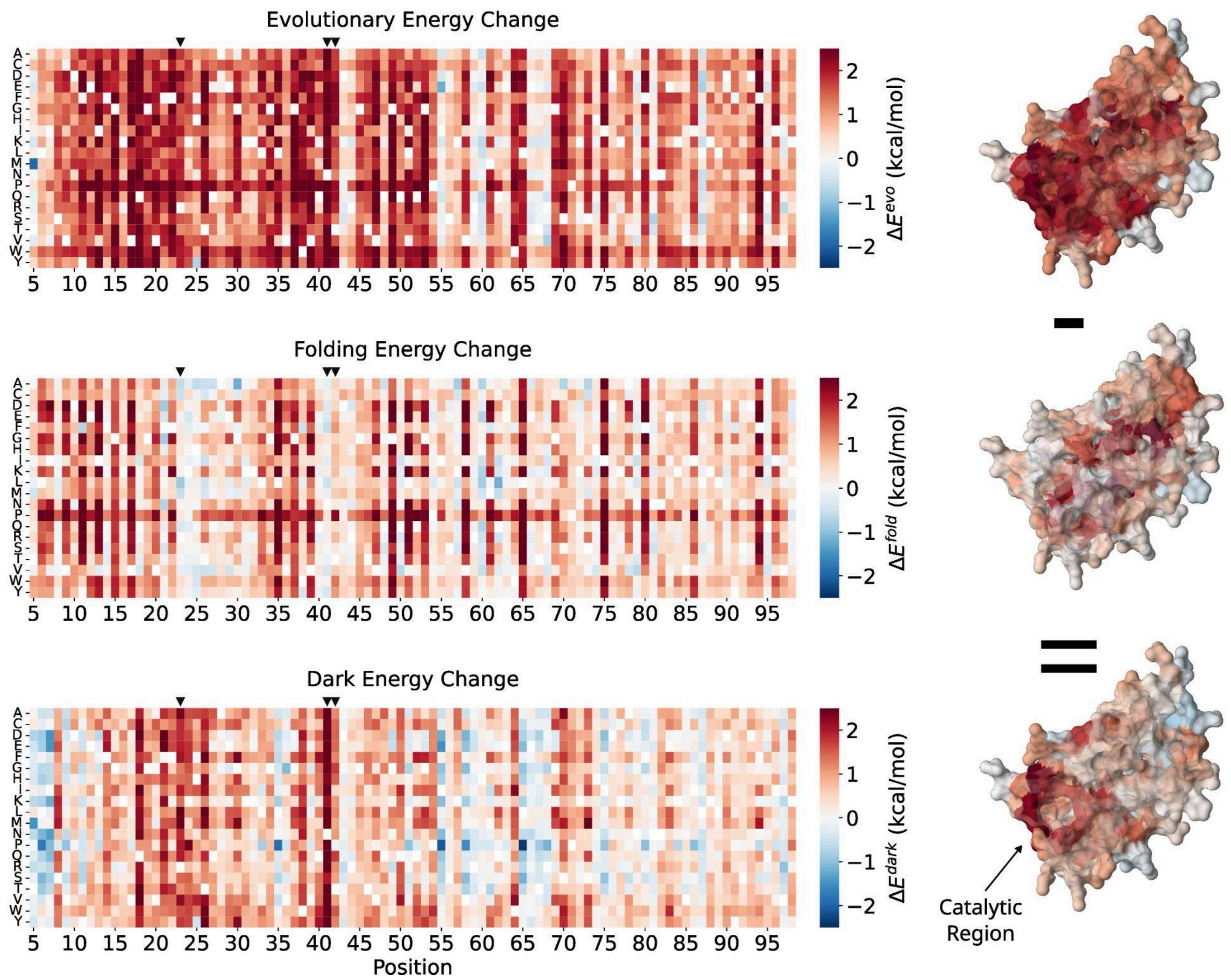

Figure 2. Three energies projected on Human Acylphosphatase-1 (ACYP1). The evolutionary energy change (top) was computed with ESM1v and rescaled with the protein selection temperature. The folding energy change (middle) was computed with AWSEM-PottsMPNN blend [25]. The dark energy (bottom) is defined as the difference between the evolutionary and the folding energy change for each single-site variant. Residues in the active site (Arg23, Asn41 and Thr42) are highlighted with an arrow. The experimental structure (PDB: 2VH7) is colored with the weighted-site average of each energy, in semitransparent surface. Annotated catalytic residues (and others close to them) present high dark energy.

**Low-throughput evidence for the cost of function in protein evolution**

Around 1990, refined site-directed mutagenesis protocols and protein stability measurements came together with existing methods to study enzymatic and binding reactions. This combination made it possible to investigate both protein stability and function, and to compare the effects of a given mutation on these two aspects of protein behavior. The interplay of function and stability was first studied on a case-by-case basis. For example, it was reported that active-site mutations that decrease the specific activity of pig citrate synthase also cause an increase in its conformational stability [25] The accumulation of similar reports led to the proposal of a trade-off between protein function and stability at protein functional sites [26–28]. However, functional site residues of some proteins do make positive contributions to both stability and function [29], which were proposed to be coupled in some scenarios [28]. For practical reasons, these classical, low-throughput studies defined catalytic/binding sites as the sets of residues in direct contact with the substrate or other target molecules. However, the well-known existence of allosterism implies that some amino acids far from the protein catalytic/binding site are also relevant for function. The interplay of stability and function for these amino acids was left unstudied at this stage.

The accumulation of experimental evidence and the development of computational methods eventually allowed a combined study of the interplay of function, stability and misfolding in tens of globular domains [28]. The integration of anecdotal evidence revealed general trends: at catalytic sites, selection for function overrules selection for stability, but there is no general anticorrelation between function and stability. At binding sites, selection for stability plays a secondary role, but is not coupled to selection for function. The definition of functional sites as those where stability rules are not fully satisfied suggested that it should be possible to develop a method that can identify these sites by analyzing the molecular interactions of a protein structure [27].

**High-throughput evidence for Dark Energy**

In the last five years, technological advances have enabled the community to explore at scale how systematic sequence perturbations affect the folding energy landscape and the corresponding evolutionary landscape. On the one hand, experimental techniques have advanced to enable the proliferation of deep mutational scanning (DMS) experiments, in which nearly every possible single-site sequence variation can be measured in the same conditions [30,31]. DMS studies have scaled from single proteins to many [32,33], and from single readouts to the measurement of multiple phenotypes, such as enzymatic activity and protein abundance [34,35]. On the other hand, deep-learning methods have revolutionized fitness and stability prediction. Trained with millions of parameters, protein language models (pLMs) can predict changes in the likelihood of any protein sequence, and therefore in its evolutionary potential, upon point mutations, without requiring sequence alignments [36,37]. Inverse-folding (IF) models, deep-learning architectures trained to generate new sequences conditioned on a given backbone, have emerged as a powerful unsupervised alternative -although also limited- for predicting stability changes [38–40]. Several research groups have implemented and leveraged these novel experimental and computational methods to systematically compare evolutionary fitness costs with the corresponding folding-stability changes, revealing the presence of a dark energy component that accounts for the cost of biological functions beyond folding. In the following, we discuss some relevant approaches and summarize them in Table 1.

Interpreting a protein abundance DMS as a stability proxy and comparing it with an activity DMS for the same protein by setting thresholds in each readout, Cagiada, Lindorff-Larsen and others developed a systematic classification of protein sites, finding that many low-activity, high-abundance positions play functional roles in PTEN and NUDT15 [41]. Replacing the experimental abundance scores with a force-field estimation of stability and the activity assay with a sequence evolutionary score, the threshold-based classification has been applied to other datasets, identifying stable-but-inactive (SBI) variants [33,42,43]. A more recent high-throughput implementation of the method, FunC-ESMs, using as stability proxy the inverse-folding ESM-IF1 and as evolutionary score the pLM ESM1b, allowed to classify the missense variants of the complete human proteome identifying a 60% as WT-like, 16% as total-activity-loss and 24% as SBI [44]. A Deep-learning method that computes the difference between evolutionary and folding-stability scores (a non-scaled dark energy) at a residue representation level surpassed the performance of the thresholding method on SBI prediction [45]. Without taking into account the differences in $T_{sel}$ or computing any other protein-specific calibration, raw evolutionary scores are not comparable between different protein families [46]. This limitation does not affect variant classification within each sequence but should be taken into account if the obtained functional signal aims to be compared across the proteome.

Protein abundance DMS have been scaled up using protein fragment complementation (AbundancePCA) and combined with independent assay of the abundance of the same protein bound to a partner (BindingPCA) [47]. Lehner and coworkers have combined this experimental 'doubledeepPCA' strategy with a thermodynamic interpretation of the assays, fitting a three-state (unfolded, unbound / folded, unbound / folded, bound) and two-state (unbound / bound) model to the data with a neural network, MoCHi [48]. The specific binding free energy obtained from the thermodynamic model reduces to dark energy when the folded-population correction is neglected [49]. Such binding and other non-folding functional energetic contributions were scored for each single-site variant of several proteins, representing 'allosteric maps' [47,50–53]. Where only the abundance DMS is available, taking it as a nonlinear proxy of the stability changes, Lehner and coworkers computed an empirical, protein-specific functional score as the residual respect to the corresponding evolutionary scores, identifying functional residues in 500 protein human protein domains [32]. This approach was extended at large scale [54,55] by replacing the abundance measurements with supervised stability predictors and spotting allostery as a widespread cause of loss-of-function variant pathogenicity in the human proteome [55].

The first explicit dark energy computations [17] were done by comparing a mega-scale folding-stability DMS dataset [33] with the corresponding evolutionary potential changes predicted with a pLM. About 25% of the positions of the folded globular proteins display some significant dark energy, beyond a fitted threshold of ~1 kcal/mol [17]. In addition, for proteins for which a stability DMS is not available yet, estimating the stability changes with the physics-based potential AWSEM [19], it was found that dark energy localizes around enzyme catalytic centers and smoothly decays as one leaves the center's vicinity. With the development of a blend of inverse-folding and physics-based models that leverages IF predictive power but reduces the impact of functional conservation conflation in stability predictions, it was possible to separate the folding and dark energy components in human missense variants, where oncogenic drivers and gain-of-function mutants exhibited particularly high dark energy [56].

| Method name | Fitness proxy | Stability proxy | Output type | Method description | First Author | Last Author | Ref |
|---|---|---|---|---|---|---|---|
| Functional Model | Activity* | Abundance* | Residue or Variant Class | Threshold-based classification | Cagiada | Lindorff-Larsen | [41] |
| | DCA | Rosetta | | | | | |
| | GEMME | | | | Høie | | [43] |
| | | | | | Cagiada | | [42] |
| FunC-ESMs | ESM1b | ESM-IF1 | | | | | [44] |
| Functional residue predictor | GEMME | ΔΔG* | | | Tsuboyama | Rocklin | [33] |
| MoCHI | DMS* | DMS* | Free energy | Neural Network fit of multistate thermodynamic models | Faure | Lehner | [48] |
| | Binding* | Abundance* | | | | | [47] |
| | | | | | Weng | | [50] |
| | | | | | Martí-Aranda | | [53] |
| | Activity* | | | | Beltran | | [51] |
| | | | | | Folgado | | [52] |
| Residual | ESM1v | | Functional Score | Residual to nonlinear fit | Beltran | | [32] |
| | | ThermoMPNN | | | Liao | | [55] |
| SPURS | ESM1v | SPURS | | | Li | Luo | [54] |
| DETANGO | ESM1v | FoldX | | pLM reprogramming | Ding | | [45] |
| | | SPURS | | | | | |
| Dark Energy | ESM2 | ΔΔG* | Free energy | Energy difference | Galpern | Ferreiro & Wolynes | [17] |
| | | AWSEM | | | | | |
| | ESM1v | AWSEM-PottsMPNN | | | | Dias & Frazer | [56] |

Table 1. High-throughput approaches quantifying the cost of protein function beyond folding combining a fitness and a stability proxy, including experimental assays (indicated with *) and computational methods.

## Conclusions and open questions

The concept of "dark energy" in protein evolution provides a powerful quantitative framework for understanding the fundamental tension between folding stability and biological function. By measuring the discrepancy between evolutionary and folding free energies, we can now assign energetic costs, measured in kcal/mol, to specific functional constraints that have shaped protein sequences throughout evolution. Initial findings reveal that approximately a quarter of sites in globular proteins are subject to significant functional constraints that extend beyond those of folding stability [17]. Dissecting folding and functional contributions to protein physiology reveals several patterns that are consistent across the different approaches. Around one half of the protein sequence variants causing loss of function can not be explained solely with their effects on folding [32,41,43]. Dark energy can help map active sites, as well as regulatory and binding sites [17,42], and allosteric interactions between them [47,50,51,55]. Dark energy generally decays with the distance to functional sites [17,51,53,55] and variations in dark energy help us understand the molecular mechanism of disease-related mutations [32,32,42], with distinctive signatures in oncogenic drivers and gain-of-function mutations [56]. Comparative mapping across homologous proteins reveals both conserved and lineage-specific energetic features, underscoring the evolutionary plasticity of functional constraints [45,53]. It is apparent that the dark energy localizes around known catalytic centers and binding interfaces and offers a natural explanation for why functional regions often exhibit higher levels of local frustration. Looking forward, key open questions remain: What determines the maximum dark energy a protein can tolerate relative to its global stability? Are certain biological functions inherently more costly than others? Can we engineer proteins with tailored dark energy profiles? Why do molecular natural contrivances like the ribosome achieve such massive scales, perhaps to distribute functional costs across many subunits? The dark energy framework clarifies the physical basis of protein evolution and provides practical tools for predicting functional sites, interpreting disease variants, and hopefully guiding protein engineering efforts.

## Acknowledgments

E.A.G. acknowledges support from the European Union's Horizon Europe under the grant agreement No 101208464. The authors acknowledge L.A.M.C. for his recent inspiring performance. Research for this publication has been partially carried out in the Barcelona Collaboratorium for Modelling and Predictive Biology. Funded by the European Union. Views and opinions expressed are however those of the author(s) only and do not necessarily reflect those of the European Union. Neither the European Union nor the granting authority can be held responsible for them. IES and DUF are supported by the Consejo de Investigaciones Científicas y Técnicas (CONICET); CONICET Grant PIP2022-2024—11220210100704CO and Universidad de Buenos Aires grant UBACyT 20020220200106BA.

**References and recommended reading**

Papers of particular interest, published within the period of review, have been highlighted as:

* of special interest

** of outstanding interest.

1. Bray, D. Protein molecules as computational elements in living cells. *Nature* **376**, 307–312 (1995).
2. Sánchez, I. E., Galpern, E. A., Garibaldi, M. M. & Ferreiro, D. U. Molecular Information Theory Meets Protein Folding. *J. Phys. Chem. B* **126**, 8655–8668 (2022).
3. Ferreiro, D. U., Komives, E. A. & Wolynes, P. G. Frustration, function and folding. *Curr. Opin. Struct. Biol.* **48**, 68–73 (2018).
4. Bryngelson, J. D. & Wolynes, P. G. Spin glasses and the statistical mechanics of protein folding. *Proc. Natl. Acad. Sci.* **84**, 7524–7528 (1987).
5. Ferreiro, D. U., Hegler, J. A., Komives, E. A. & Wolynes, P. G. Localizing frustration in native proteins and protein assemblies. *Proc. Natl. Acad. Sci. U. S. A.* **104**, 19819–19824 (2007).
6. Parra, R. G., Komives, E. A., Wolynes, P. G. & Ferreiro, D. U. Frustration in physiology and molecular medicine. *Mol. Aspects Med.* **103**, 101362 (2025).
7. Parra, R. G. & Ferreiro, D. U. Frustration, dynamics, and catalysis. *Curr. Opin. Struct. Biol.* **94**, 103127 (2025).
8. Freiberger, M. I. *et al.* Local energetic frustration conservation in protein families and superfamilies. *Nat. Commun.* **14**, 8379 (2023).
9. Cao, H. *et al.* SF-Cluster: Frustration-Guided MSA Subsampling for Alternative Protein Conformation Recovery. Preprint at https://doi.org/10.48550/arXiv.2607.00180 (2026).
10. Maynard Smith, J. Natural Selection and the Concept of a Protein Space. *Nature* **225**, 563–564 (1970).

11. Wolynes, P. G. Evolution, energy landscapes and the paradoxes of protein folding. *Biochimie* **119**, 218–230 (2015).

12. Shakhnovich, E. I. & Gutin, A. M. Implications of thermodynamics of protein folding for evolution of primary sequences. *Nature* **346**, 773–775 (1990).

13. Pande, V. S., Grosberg, A. Y. & Tanaka, T. Statistical mechanics of simple models of protein folding and design. *Biophys. J.* **73**, 3192–3210 (1997).

14. Lindquist, S. Protein Folding Sculpting Evolutionary Change. *Cold Spring Harb. Symp. Quant. Biol.* **74**, 103–108 (2009).

15. Figliuzzi, M., Jacquier, H., Schug, A., Tenaillon, O. & Weigt, M. Coevolutionary Landscape Inference and the Context-Dependence of Mutations in Beta-Lactamase TEM-1. *Mol. Biol. Evol.* **33**, 268–280 (2016).

16. Possenti, A., Vendruscolo, M., Camilloni, C. & Tiana, G. A method for partitioning the information contained in a protein sequence between its structure and function. *Proteins Struct. Funct. Bioinforma.* **86**, 956–964 (2018).

17. ** Galpern, E. A., Bueno, C., Sánchez, I. E., Wolynes, P. G. & Ferreiro, D. U. Probing the dark energy in the functional protein universe. *Proc. Natl. Acad. Sci.* **123**, e2531111123 (2026).

This paper introduces and defines dark energy as the gap between the evolutionary free energy of sequences and the physical folding free energy, quantifying the cost of function beyond folding. In addition, using experimental data and computational methods, dark energy is probed throughout the protein universe, finding that it is significant in about 25% of sites and concentrates around catalytic sites in enzymes. It is shown also that the relationship of dark energy and a functional physical free energy, such as the binding energy to a partner, can be used to define a functional selection temperature.

18. Morcos, F., Schafer, N. P., Cheng, R. R., Onuchic, J. N. & Wolynes, P. G. Coevolutionary information, protein folding landscapes, and the thermodynamics of natural selection. *Proc. Natl. Acad. Sci.* **111**, 12408–12413 (2014).

19. Davtyan, A. *et al.* AWSEM-MD: Protein structure prediction using coarse-grained physical potentials and bioinformatically based local structure biasing. *J. Phys. Chem. B* **116**, 8494–8503 (2012).

20. Morcos, F. *et al.* Direct-coupling analysis of residue coevolution captures native contacts across many protein families. *Proc. Natl. Acad. Sci.* **108**, E1293–E1301 (2011).

21. Frazer, J. *et al.* Disease variant prediction with deep generative models of evolutionary data. *Nature* **599**, 91–95 (2021).

22. Galpern, E. A., Roman, E. A. & Ferreiro, D. U. Inferring protein folding mechanisms from natural sequence diversity. *Biophys. J.* https://doi.org/10.1016/j.bpj.2025.06.034 doi:10.1016/j.bpj.2025.06.034.

23. Miyazawa, S. Selection originating from protein stability/foldability: Relationships between protein folding free energy, sequence ensemble, and fitness. *J. Theor. Biol.* **433**, 21–38 (2017).

24. Galpern, E. A., Caamaño, F. & Ferreiro, D. U. Predicting Protein Folding Dynamics Using Sequence Information. in *Protein Evolution* (eds Khan, S. M. & Pazos, F.) vol. 2979 17–31 (Springer US, New York, NY, 2026).

25. Zhi, W., Srere, P. A. & Evans, C. T. Conformational stability of pig citrate synthase and some active-site mutants. *Biochemistry* **30**, 9281–9286 (1991).

26. Schreiber, G., Buckle, A. M. & Fersht, A. R. Stability and function: two constraints in the evolution of barstar and other proteins. *Structure* **2**, 945–951 (1994).

27. Beadle, B. M. & Shoichet, B. K. Structural Bases of Stability–function Tradeoffs in Enzymes. *J. Mol. Biol.* **321**, 285–296 (2002).

28. Sánchez, I. E., Tejero, J., Gómez-Moreno, C., Medina, M. & Serrano, L. Point Mutations in Protein Globular Domains: Contributions from Function, Stability and Misfolding. *J. Mol. Biol.* **363**, 422–432 (2006).

29. Kragelund, B. B. *et al.* Conserved Residues and Their Role in the Structure, Function, and Stability of Acyl-Coenzyme A Binding Protein. *Biochemistry* **38**, 2386–2394 (1999).

30. Rubin, A. F. *et al.* MaveDB 2024: a curated community database with over seven million variant effects from multiplexed functional assays. *Genome Biol.* **26**, 13 (2025).

31. Notin, P. *et al.* ProteinGym: Large-Scale Benchmarks for Protein Fitness Prediction and Design. in *Advances in Neural Information Processing Systems* vol. 36 64331–64379 (2023).

32. ** Beltran, A., Jiang, X., Shen, Y. & Lehner, B. Site-saturation mutagenesis of 500 human protein domains. *Nature* **637**, 441–449 (2025).

The authors measured the protein abundance at site-saturation for 500 domains and compared it with the corresponding ESM1v log-odds, fitting a sigmoid. The residuals to these fits identify mutations with larger or smaller effects on evolutionary fitness than can be accounted for by changes in stability.

33. Tsuboyama, K. *et al.* Mega-scale experimental analysis of protein folding stability in biology and design. *Nature* **620**, 434–444 (2023).

34. Vanella, R. *et al.* Understanding activity-stability tradeoffs in biocatalysts by enzyme proximity sequencing. *Nat. Commun.* **15**, 1807 (2024).

35. Simon, J. J., Fowler, D. M. & Maly, D. J. Multiplexed profiling of intracellular protein abundance, activity, interactions and druggability with LABEL-seq. *Nat. Methods* **21**, 2094–2106 (2024).

36. Hou, C., Liu, D., Zafar, A. & Shen, Y. Understanding language model scaling for protein fitness prediction. *Nat. Comput. Sci.* **6**, 778–788 (2026).

37. Bjerregaard, A., Groth, P. M., Hauberg, S., Krogh, A. & Boomsma, W. Foundation models of protein sequences: A brief overview. *Curr. Opin. Struct. Biol.* **91**, 103004 (2025).

38. Frellsen, J. *et al.* Zero-shot protein stability prediction by inverse folding models: a free energy interpretation. in *Advances in Neural Information Processing Systems* (eds Belgrave, D. et al.) vol. 38 84641–84667 (Curran Associates, Inc., 2025).

39. Dutton, O. *et al.* Improving Inverse Folding models at Protein Stability Prediction without additional Training or Data. Preprint at https://doi.org/10.1101/2024.06.15.599145 (2024).

40. Birnbaum, F. & Keating, A. E. Beyond native sequence recovery: Improved modeling of the sequence-energy landscape of protein structures. *Proc. Natl. Acad. Sci.* **123**, e2535494123 (2026).

41. Cagiada, M. *et al.* Understanding the Origins of Loss of Protein Function by Analyzing the Effects of Thousands of Variants on Activity and Abundance. *Mol. Biol. Evol.* **38**, 3235–3246 (2021).

42. Cagiada, M. *et al.* Discovering functionally important sites in proteins. *Nat. Commun.* **14**, 4175 (2023).

43. Høie, M. H., Cagiada, M., Beck Frederiksen, A. H., Stein, A. & Lindorff-Larsen, K. Predicting and interpreting large-scale mutagenesis data using analyses of protein stability and conservation. *Cell Rep.* **38**, 110207 (2022).

44. ** Cagiada, M., Jonsson, N. & Lindorff-Larsen, K. Decoding molecular mechanisms for loss-of-function variants in the human proteome. Preprint at https://doi.org/10.7554/eLife.108160.1 (2025).

This work presents FunC-ESMs, a functional classifier model and applies it to the human-proteome. FunC-ESMs separates ‘WT-like’ variants from deleterious variants imposing a threshold on their ESM-1b (a pLM) score. For those variants identified as deleterious, the a threshold in the inverse-folding model ESM-IF log-odds is then used to assign the ‘total-loss’ label to those that cause structural destabilisation or the ‘stable-but-inactive’ label to those that affect function but not stability.

45. * Ding, K., Li, Z., Tu, T., Luo, J. & Luo, Y. Deconvolving mutation effects on protein stability and function with disentangled protein language models. Preprint at https://doi.org/10.64898/2026.02.03.703560 (2026).

Introduces DETANGO, a deep-learning method that computes the difference between the pLM evolutionary signal and the folding-stability changes at the level of the pLM embeddings. This operation reprograms the pLM to output a functional plausibility score for each single-point mutation.

46. Orenbuch, R. *et al.* Proteome-wide model for human disease genetics. *Nat. Genet.* **57**, 3165–3174 (2025).

47. Faure, A. J. *et al.* Mapping the energetic and allosteric landscapes of protein binding domains. *Nature* **604**, 175–183 (2022).

48. * Faure, A. J. & Lehner, B. MoCHI: neural networks to fit interpretable models and quantify energies, energetic couplings, epistasis, and allostery from deep mutational scanning data. *Genome Biol.* **25**, 303 (2024).

Introduces MoCHI, a deep-learning framework and software package for fitting two- and three-state equilibria thermodynamic models to deep mutational scanning data. The method can be used to obtain the binding and folding free energy changes.

49. Otwinowski, J. Biophysical Inference of Epistasis and the Effects of Mutations on Protein Stability and Function. *Mol. Biol. Evol.* **35**, 2345–2354 (2018).

50. * Weng, C., Faure, A. J., Escobedo, A. & Lehner, B. The energetic and allosteric landscape for KRAS inhibition. *Nature* **626**, 643–652 (2024).

In this article multiple global atlases of inhibitory allosteric communication in KRAS are presented, quantifying the impact of >26,000 mutations on the folding of KRAS and its binding to six interaction partners. The authors performed BindingPCA and AbundancePCA experiments and imputed the data to the MoCHI framework to obtain the folding free energy and the specific binding free energies for each partner, obtaining the allosteric maps.

51. Beltran, A., Naqvi, M. M., Faure, A. J. & Lehner, B. The allosteric landscape of the Src kinase. *Sci. Adv.* **12**, eaea2726 (2026).

52. Folgado, C., Beltran, A. & Lehner, B. Conservation and divergence in the allosteric architectures of five human protein kinases. Preprint at https://doi.org/10.64898/2026.08.04.742685 (2026).

53. Martí-Aranda, A. & Lehner, B. Seven complete comparative maps of allosteric mutations in a protein family. *Nat. Commun.* **17**, 5487 (2026).

54. Li, Z. & Luo, Y. Generalizable and scalable protein stability prediction with rewired protein generative models. *Nat. Commun.* **17**, 891 (2025).

55. ***** Liao, X. & Lehner, B. Allostery is a widespread cause of loss-of-function variant pathogenicity. *Nat. Commun.* (2026).

This paper presents large-scale experimental measurements and deep-learning models to provide evidence that allostery is a widespread cause of loss-of-function variant pathogenicity in human genetic diseases. The authors compute the residuals to a nonlinear (LOESS) fit to separate folding-stability from ThermoMPNN from the evolutionary scores computed with ESM1v.

56. ***** Galpern, E. A. *et al.* Blending physics-based and inverse folding models to disentangle variant effects on stability and function. Preprint at https://doi.org/10.64898/2026.07.30.741764 (2026).